\documentclass[preprint,12pt]{elsarticle}
\usepackage{amsmath}    
\usepackage{amssymb}
\usepackage{natbib}

\usepackage{graphicx}
\usepackage{epstopdf}
\usepackage{dcolumn}
\usepackage{bm}
\usepackage{float}
\usepackage{breqn}

\biboptions{comma,square}
\journal{Commun Nonlinear Sci Numer Simulat}

\begin{document}
\begin{frontmatter}
\title{Detecting Directed Interactions of Networks by Random Variable Resetting}
\author{Rundong Shi, Changbao Deng,  Shihong Wang$^{1\ast}$}

\address{School of Sciences, Beijing University of Posts and Telecommunications,Beijing 100876,China\\
$^{\ast}$Corresponding author shwang@bupt.edu.cn}
\date{\today}
\maketitle

\begin{abstract}
We propose a novel method of detecting directed interactions of a general dynamic network from measured data.
By repeating random state variable resetting of a target node and appropriately averaging over the measurable data,
the pairwise coupling function between the target and the response nodes can be inferred.
This method is applicable to a wide class of networks with nonlinear dynamics, hidden variables and strong noise.
The numerical results have fully verified the validity of the theoretical derivation.
\end{abstract}


\end{frontmatter}

Complex networks are investigated in many scientific areas. Due to collective behavior and functional diversity \cite{1,2,3,4}, network reconstruction from measured data, especially inferring network interaction is one of the most challenging topics.
Investigation of the network structure helps in understanding how it works, however, there are no easy way of direct detection.
For instance, it is quite difficult to directly detect the interactions among different brain regions which are only examined through functional connectivity analysis of functional magnetic resonance imaging (fMRI) \cite{5,6}.

The existing reconstruction methods propose to detect the dynamics of complex systems \cite{7,8,9,10,11,12,13,14,15} or to detect network connectivity \cite{16,17,18,19,20,21,22,23}.
Controlling, as a proactive approach, is usually adopted in real systems, such as synchronization and desynchronization controlling. Adopting a proactive controlling approach may reveal the entire topology, such as driving-response controlling \cite{3}, copy-synchronization \cite{24} and random phase resetting \cite{25}.
Random phase resetting method is to reconstruct the topology and interaction functions of a general oscillator network by repeatedly reinitializing the phases of all oscillators.

{Understanding the mechanism of the interaction between nodes is key to understand how a network works}.
In this Letter, we propose a novel method of reconstructing the directed interactions of a general dynamical network.
Our idea is to directly reinitialize the state variable of  {one node (called target node)}, through the state variable of the response nodes to acquire
the coupling functions of the target node to the others.
This method is applicable to a wide class of networks with nonlinear dynamics, hidden variables and strong noise.

Let us generalize the discussion to networks of interacting systems with pairwise interaction.
We consider
\begin{eqnarray}
\dot{x}_{i}(t) &=&f_{i}(x_{i})+\sum_{j=1}^{N}h_{ij}(x_{i},x_{j},{u}_{ij})+\eta_{i}(t), \\
\dot{u}_{ij}(t) &=&g(u_{ij})
\end{eqnarray}
where $f_{i}$ describes the intrinsic dynamics of node $i$, $h_{ij}$ denotes the interaction function of node $j$ to node $i$, and $u_{ij}$ is a hidden variable.
Here we assume white noise $\eta_{i}$ with zero mean and the following statistics $\langle\eta_{i}(t)\rangle = 0$, $\langle\eta_{i}(t)\eta_{j}(t')
\rangle = \sigma_{i}^{2}\delta_{ij}\delta(t-t')$.

Our aim is to infer how $x_{j}$ influences $x_{i}$, $i=1,2,...,N, \ i\neq j$, when there exist disturbances and unmeasurable variables (hidden variables) in the system.
We rewrite Eq.(1) as
\begin{equation}
\dot{x}_{i}(t) =f_{i}(x_{i})+h_{ij}(x_{i},x_{j},{u}_{ij})+\sum_{k\neq j}h_{ik}(x_{i},x_{k},{u}_{ik})+\eta_{i}(t). \\
\end{equation}

The approach is based on the following assumption: we can arbitrarily reinitialize the state variable $x_{j}$ $m$ times, where $m\gg 1$.
Now we introduce the core idea of the approach.
 For each reinitialization moment, we rewrite Eq.(3) as
\begin{equation}
\dot{x}_{i}(t) =\bar{h}_{ij}(x_{j})+r_{ij},
\end{equation}
where $\bar{h}_{ij}$ denote a mean effect of variable $x_{j}$ on variable $x_{i}$ for $m$ times through randomly resetting $x_{j}$, and $r_{ij}$ are their
fluctuations. In Eq.(4), $\bar{h}_{ij}$ can be described through averaging the right hand side of Eq.(3) as

\begin{dmath}
\bar{h}_{ij}(x_{j})=\langle h_{ij}(x_{i},x_{j},{u}_{ij})\rangle +\langle f_{i}(x_{i})\rangle +\langle \sum_{k\neq j}h_{ik}(x_{i},x_{k},{u}_{ik}) \rangle +\langle \eta_{i}(t)\rangle
\end{dmath}

In the right hand side of Eq.(5), the first item depends on the variable $x_{j}$ and directly represents the coupling function from $j$ to $i$, while it reflects the average effect of the hidden variable $u_{ij}$. The second and the third items represent the average effect of the local dynamics and other nodes on $x_{i}$, and the fourth item $\langle \eta_{i}(t)\rangle \approx 0$.

In Eq. (4), the fluctuations $r_{ij}$ have the following statistical characteristics
\begin{equation}
\langle r_{ij}\rangle \approx 0
\end{equation}
Furthermore, arbitrarily reinitializing $x_{j}$ results in no dependence of $r_{ij}$ on $x_{j}$, thus given any function $F(x_{j})$, we have the following statistical results
\begin{equation}
\langle r_{ij}F(x_{j})\rangle = \langle r_{ij}\rangle  \langle F(x_{j})\rangle \approx 0
\end{equation}

Considering the analysis above, Eq.(4) denotes a directed interaction function of node $j$ to node $i$, where the fluctuations $r_{ij}$ are independent of the variable $x_{j}$.
We name $\bar{h}_{ij}$ the equivalent coupling function of $j$ to $i$.
Now our task is to depict $\bar{h}_{ij}(x_{j})$ from measurable data ensembles.

We introduce the computation process.
HOCC and VELSA methods \cite{12,13,14} can be used to solve Eq.(4), here we use HOCC method \cite{13}.
Randomly reinitialize variable $x_{j}(t)$ with a resetting time interval
$\tau_{reset}$ and acquire all variables $x_{i}$ with measurement time interval $\tau_{measure}$, and further calculate $\dot{x}_{i}(t)=\frac{x_{i}(t+\tau_{measure})-x_{i}(t)}{\tau_{measure}}, i=1,2,3,...,N, i\neq j$. We prepare the following data ensembles $\dot{x}_{i}(t_{1}),\dot{x}_{i}(t_{2}),...,\dot{x}_{i}(t_{m})$ and $x_{j}(t_{1}),x_{j}(t_{2}),...,x_{j}(t_{m})$.

To solve Eq. (4), we assume that $\bar{h}_{ij}(x_{j})$ can be generally expanded by a
basis set as
\begin{equation}
\bar{h}_{ij}(x_{j}) \approx
\sum_{k=1}^{n}\bar{A}_{ij,k}L_{k}(x_{j})
\end{equation}
Defining $\bar{\textbf{A}}_{ij}=[\bar{A}_{ij,1}, \bar{A}_{ij,2}, ..., \bar{A}_{ij,n}]$, we rewrite Eq. (4) as
\begin{equation}
\dot{x}_{i} = \bar{\textbf{A}}_{ij}\textbf{L}(x_{j})+r_{ij}
\end{equation}
Due to $\langle r_{ij} \textbf{L}^{T}(x_{j})\rangle \approx \textbf{0} $, $\bar{\textbf{A}}_{ij}$ is calculated as the form
\begin{equation}
\bar{\textbf{A}}_{ij} = \langle\dot{x}_{i} \textbf{L}^{T}
(x_{j})\rangle \langle \textbf{L}(x_{j})
\textbf{L}^{T}(x_{j})\rangle ^{-1}
\end{equation}
where $T$ denotes the transpose of matrix of sampling data.
In Eq.(10), $\dot{x}_{i}$ can be computed from measurable ensembles, and $\textbf{L}(x_{j})$ can be taken by using self-consistent method \cite{13} and $\langle \textbf{L}(x_{j})\textbf{L}^{T}(x_{j})\rangle$ can be computed from data ensembles $x_{j}(t_{1})$, $x_{j}(t_{2})$, $...$, $x_{j}(t_{m})$.

Our theoretical findings have the following characteristics:
(i) Only through arbitrarily resetting the variable of the target node, our reconstruction method can infer the directed coupling functions of the target node to the others from measured data.
(ii)  {To deduce the coupling function between two nodes, the data ensembles is the key. }If someone foreknows the form of the coupling function by using some methods, he or she needs to determine the parameters of the coupling function, otherwise, the coupling function is generally expanded by a basis set.
(iii) If randomly resetting the variables of all the nodes in a network, this approach can infer the interactions of the entire network.
(iv) In actual complex systems, some variables cannot be  {quantified}, but they affect the states of nodes and the interactions between nodes.
Our proposed method is applicable to the networks with hidden variables and nodes, even under a strong influence of white noise.
In our theoretical findings, the hidden variable $u_{ij}$ in Eq.(3), can be transformed to the average effect and embedded in the coefficients of coupling function $\bar{h}_{ij}(x_j)$, and these coefficients essentially or indirectly denote the weights of coupling.

 {A real neural network is a high-dimensional and complex system. Only low-dimensional data can be easily measured, such as the membrane potential of neurons. In addition, due to the nonlinearity of real neural networks and high noise, detecting the interactions of neural networks is challenging}. In this Letter, we apply our theoretical method in neural networks and verify our findings through numerical simulations.

First consider a neural network with $N=30$ Hodgkin-Huxley neurons \cite{25} shown in Fig.1, which is produced by Morita's method \cite{22,26}. The dynamics of node $i$ can be expressed as:
\begin{dmath}
C_{m,i}\dot{V}_{i}=-g_{L,i}(V_{i}-E_{L})-g_{Na,i}m_{i}^{3}h_{i}(V_{i}-E_{Na})-g_{K,i}n_{i}^{4}(V_{i}-E_{K})+I_{inject,i}+I_{i}^{(syn)}
\end{dmath}
where synaptic current $I_{i}^{(syn)}$ denotes the interaction function of neurons. For verifying our theoretical method, we take two kinds
of synapses, electrical and chemical synapses, i.e., two kinds of coupling functions, and $I_{i}^{(syn)}$ is expressed by the form
\begin{dmath}
I_{i}^{(syn)}= \sum_{j\neq i}g_{ij}(V_{j}-V_{i})+\sum_{j\neq i}\gamma_{ij}\frac{V_{rev}-V_{i}}{1+\exp(-(V_{j}-V_{th})/\sigma)}
\end{dmath}
where $g_{ij}$ and $\gamma_{ij}$ are the elements of the adjacency matrices $\textbf{G}$ and $\bm{\Gamma}$, respectively.
$\textbf{G}$ is a symmetric matrix standing for the interactions of electrical synapses while $\bm{\Gamma}$ stands for the interactions
of chemical synapses and is asymmetric.
When the dynamics of node $i$ is affected by node $j$ via the electrical synapses or the chemical synapses, $g_{ij}$ and $\gamma_{ij}$  {are distributed} uniformly in $(0, 1)$ expressed by $U(0,1)$
whereas $g_{ij}=0$, $\gamma_{ij}=0$ for nonlinks.
$V_{rev}=110 \ mV$, $V_{th}=100 \ mV$ and $\sigma=0.01$.

\begin{figure}[H]
\includegraphics[width=14cm]{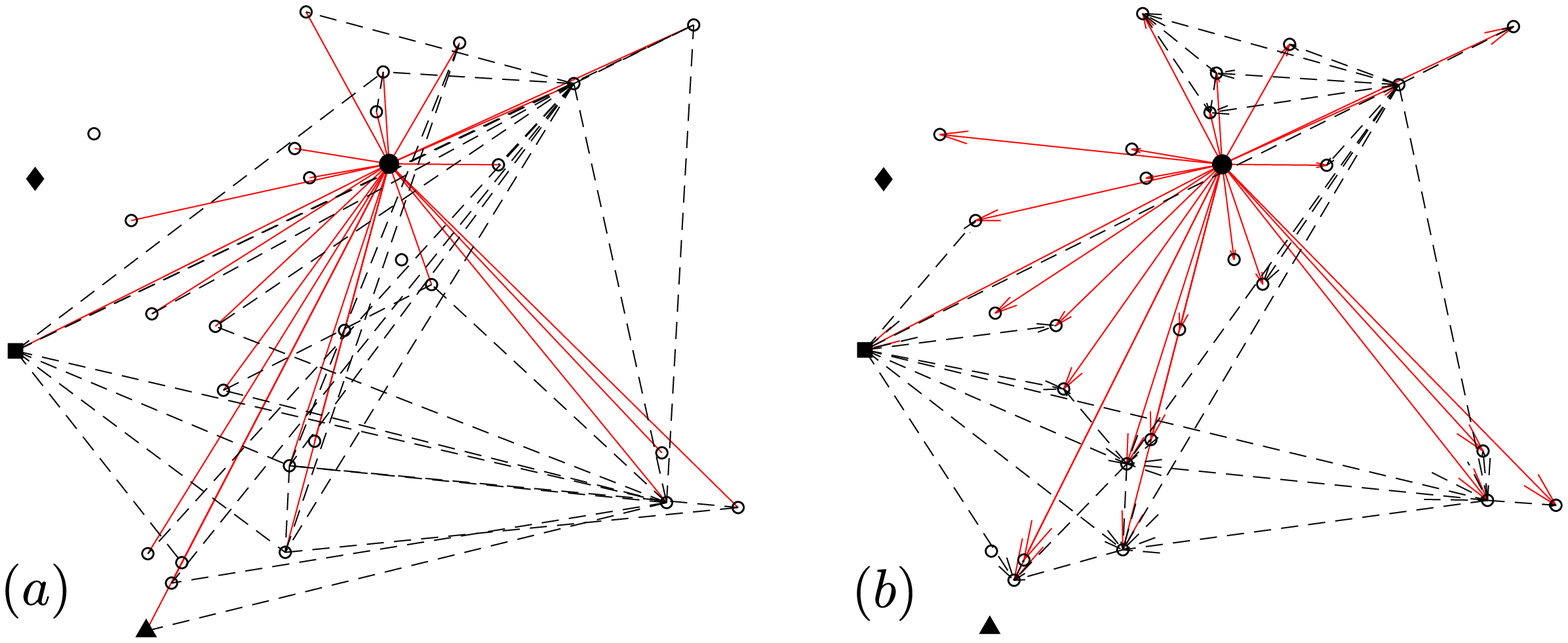}
\caption{\label{fig:epsart} Thirty-node neuron network used for illustrating our method. The symbols $\bullet$,
\footnotesize$\blacksquare$ \small, $\blacktriangle$ and $\blacklozenge$ represent node 1, node 3, node 26 and node 29, respectively. Red solid lines denote
the links of node 1 to other twenty-nine nodes, and black dashed lines denote other links. (a) Electrical synapse network.
(b) Chemical synapse network. The direction of interaction is denoted by that of the arrow.}
\end{figure}

In Eq.(11), all parameters  {are distributed} uniformly as the followings,
$C_{m,i}\in [0.9, 1.1]\ \mu F/cm^{2}$, $g_{L,i}\in [0.25,0.35] \
ms/cm^{2}$, $g_{Na,i}\in [110,130]\ ms/cm^{2}$, $g_{K,i}\in
[30,40]\ ms/cm^{2}$, $E_{L}\in [-11.1, -10.1]\ mV$, $E_{Na}\in
[110,130]\ mV$, $E_{K}\in [-30,-20]\ mV$, and $I_{inject,i}\in
[5,15]\ mA$.

In Eq. (11) all  {gating} variables can be written as
\begin{eqnarray}
 \dot{m}_{i}&=&
 \alpha_{m,i}(V)(1-m_{i})-\beta_{m,i}(V)m_{i} \nonumber \\
 \dot{h}_{i}&=&
 \alpha_{h,i}(V)(1-h_{i})-\beta_{h,i}(V)h_{i} \\
 \dot{n}_{i}&=&
 \alpha_{n,i}(V)(1-n_{i})-\beta_{n,i}(V)n_{i} \nonumber
\end{eqnarray}
where $\alpha_{m,i}=\frac{2.5-0.1V}{e^{2.5-0.1V}-1}$,
$\beta_{m,i}=4e^{-V/18}$, $\alpha_{h,i}=0.07e^{-V/20}$,
$\beta_{h,i}=\frac{1}{e^{3-0.1V}+1}$,
$\alpha_{n,i}=\frac{0.1-0.01V}{e^{1-0.1V}-1}$,
$\beta_{n,i}=0.125e^{-V/80}$.

In an ideal situation, we can arbitrarily change the membrane potential.
Now our task is to uniformly reinitialize variable $V_{1}(t)\in [-30,\ 110] \ mV$ with a resetting time interval
$\tau_{reset}$ and to acquire all variables $V_{i}, i=2,3,...,N$,
with measurement time interval $\tau_{measure}$, and to further
reconstruct the coupling functions. In addition, all variables
$m_i$, $h_i$ and $n_i$ in Eq. (13), $i=1,2,...,30$, cannot be measured, i.e,
they are hidden variables.  {To detect the coupling functions, we
generally assume that they can be expanded by a basis set, for example, power series are chosen as a basis set.
However, due to the random state resetting of our method, we easily
acquire the statistical results of all measurable ensembles that
indicate the varied forms of coupling functions. Our aim is to determine the parameters of the coupling functions. In the simulations, the statistical results ($\dot{V}_i$, $i=2,3,...,N$) show no dependence on variable $V_1$, linear dependence, nonlinear dependence (sigmoidal function), or their summing.
Examples are shown in Fig.2.} In
Fig.2, we present the numerical values of $\dot{V}_{3}$, $\dot{V}_{26}$ and $\dot{V}_{29}$ with variable $V_{1}$ for $m=10^{5}$ times, the averages of $\dot{V}_{3}$, $\dot{V}_{26}$ and $\dot{V}_{29}$ and their actual values, respectively.
The average curve in Fig.2(f) has no sensitive dependence on $V_{1}$, which coincides with the actual one (red solid line).
The average curve in Fig.2(e) has linear dependence on $V_{1}$
whereas one in Fig.2(d) demonstrates the following feature: a
linear interval and a nonlinear interval (sigmoidal function). The
statistical estimates of coupling functions in Figs.2(d-f) display a good
agreement for no links, electrical and chemical synapses.

Based on the statistical averages,  {conclusions can be drawn on }that the coupling function of the system is expressed as the following fitting function
\begin{equation}
\bar{I}_{i1}^{(syn)}= \hat{g}_{i1}V_{1}+\hat{\gamma}_{i1}\frac{1}{1+\exp(-(V_{1}-V_{th})/
\sigma)}
\end{equation}
where $V_{th}=100 \ mV$ and $\sigma=0.01$ (Taking $\sigma=0.1$ and
$\sigma=0.001$ does not really affect the numerical results).

\begin{figure}[htp]
\includegraphics[width=14cm]{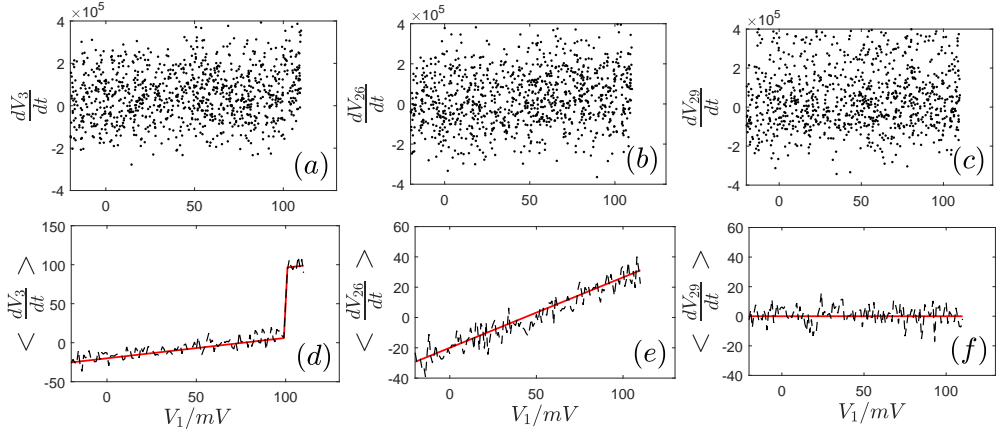}
\caption{\label{fig:epsart} Scatter plots of numerical
$\dot{V}_{3}$ (a),  $\dot{V}_{26}$ (b) and $\dot{V}_{29}$ (c) vs. the initial value $V_{1}\in [-30,\ 110]\ mV$ chosen uniformly for $m=10^{5}$.
The curves in (d-f) show the average of $\dot{V}_{3}$, $\dot{V}_{26}$ and
$\dot{V}_{29}$ vs. $V_{1}$ and their actual values (red solid lines), respectively.}
\end{figure}

We take $\tau_{reset}=100 \ ms$, $\tau_{measure}=1 \ ms$ and
compute $\hat{g}_{ij}$ and $\hat{\gamma}_{ij}$ from an ensemble of $m=10^{5}$ via Eq.(10).
In Figs. 3(a) and 3(b), we compare the numerical
$\hat{g}_{i1}$ and $\hat{\gamma}_{i1}$ with the actual values
$g_{i1}$ and $\gamma_{i1}$, respectively. They all display a very
good agreement. If we arbitrarily reinitialize all variables
$V_{j}, j=1,2,...,N$, we can estimate all forms of coupling
functions, find all interaction parameters, and further reconstruct
the coupling functions of the entire network.

To show the average error between the calculated values and the actual values,
we calculate the root mean square error,
\begin{equation}
E_{rms}=\sqrt{\frac{\sum_{i=1}^{N}(\hat{v}_{i}-v_{i})^{2}}{N}}.
\end{equation}
where $\hat{v}_{i}$ and $v_{i}$ represent the simulation results and the
corresponding actual values.
Now we discuss how the properties of reinitialization influence the final precision of the network reconstruction. Consider the number of random variable resetting $m$, the randomicity of reinitialization $K'$ and the system size $N$.
Define variable resetting with $V_{1}(t)=110-K_{1} \in [-30,\ 110] \ mV$, where $K_{1}$  {is distributed} uniformly in $U(0,K')$.
Fig.3(c) and 3(d) show the plots of $E_{rms}$ vs. $m$ and $E_{rms}$ vs. $K'$ for electrical and chemical synapses. In Fig.3(c), the results show that the reconstruction errors for electrical and chemical synapses are approximately proportional to $\frac{1}{\sqrt{m}}$. In Fig.3(d), the precision of the network reconstruction for electrical synapses increases with the strong randomicity whereas the reconstruction for chemical synapses has a high precision at $K' > 30 \ mV $. This is because there approximately exist only two kinds of dynamical states for chemical synapses, spiking behavior and resting state.

 {To further determine the dependence of $m$ and $K'$ to the reconstruction precisions, the Pearson correlation between the actual parameters $v_i$ and the inferred parameters $\hat{v}_i$ is calculated by the following form
\begin{equation}
P=\frac{\sum_{i=1}^N(v_i-\bar{v}_i)(\hat{v}_i-\bar{\hat{v}}_i)}{\sqrt{\sum_{i=1}^N(v_i-\bar{v}_i)^2 }\sqrt{\sum_{i=1}^N(\hat{v}_i-\bar{\hat{v}}_i)^2}}
\end{equation}
Figures 3(e) and (f) show the plots of $P$ vs. $m$ and $P$ vs. $K'$ for electrical and chemical synapses. In Figs.3(e-f) we know that the smallest usable $m$ is about $10^3$ for chemical and electrical synapses, and the smallest $K'$ is about $20$ for chemical synapse and $30$ for electrical synapse.}

\begin{figure}[htp]
\includegraphics[width=14cm]{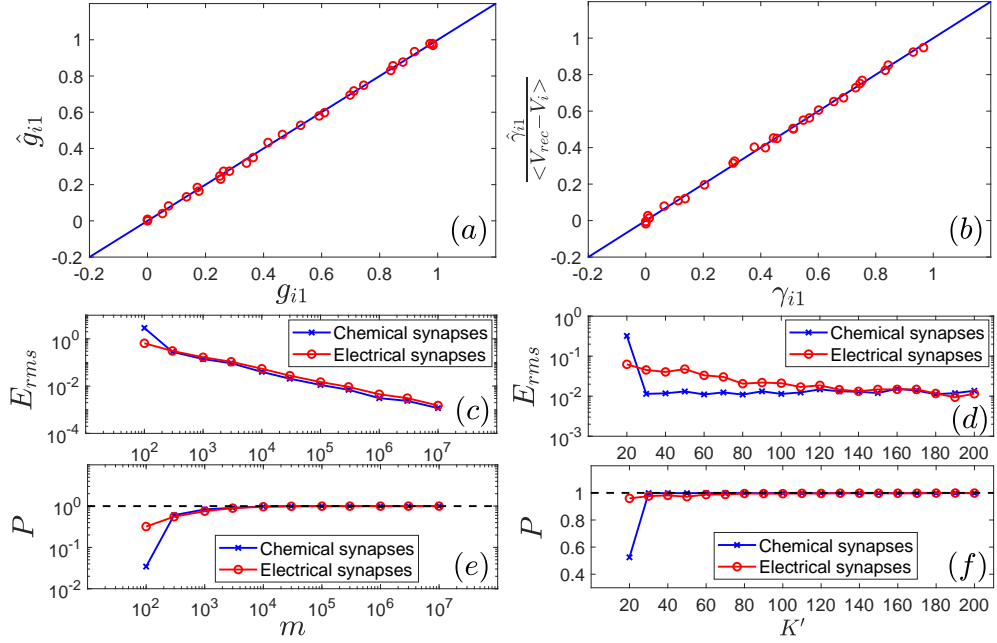}
\caption{\label{fig:epsart} Reconstruction of network of Fig.1.
(a) Numerical results $\hat{g}_{i1}$ plotted against the actual coefficients $g_{i1}$.
(b) Numerical results $\frac{\hat{\gamma}_{i1}}{\langle V_{rev}-V_{i}\rangle}$ plotted against the actual coefficients $\gamma_{i1}$.
 {(c)(e) Plots of $E_{rms}$ vs. $m$ and $P$ vs. $m$ for electrical and chemical synapses. $K'=140 \ mV $.
(d)(f) Plots of $E_{rms}$ vs. $K'$ and $P$ vs. $K'$ for electrical and chemical synapses. $m=10^{5}$.}}
\end{figure}

 {To further understand how the reconstruction precision varies with the network size $N$, two kinds of networks, fully connected networks (FCNs) and sparse networks (SNs), are chosen for the study. Each node in the FCNs is connected to all the other nodes, and each node in the SNs is connected to five other nodes. The connection strength is distributed uniformly in (0,1). We calculate $E_{rms}$ via Eq. (15) and $P$ via Eq. (16) and the computed results are plotted in Fig.4. The plots of Fig.4(c)(d) show that the reconstruction precisions of FCNs and SNs decrease slightly with the increase of the network size, and the decrease of FCNs is a little stronger than that of SNs. It can be easily understood, if a network becomes large and its connection is relatively dense, the fluctuations of $\langle \sum_{k\neq j}h_{ik}(x_{i},x_{k},{u}_{ik}) \rangle$ in Eq. (5) will increase, i.e., increasing the fluctuations $r_{ij}$ in Eq. (4). The increases of $r_{ij}$ slightly affect the reconstruction precisions.}

\begin{figure}[H]
\includegraphics[width=14cm]{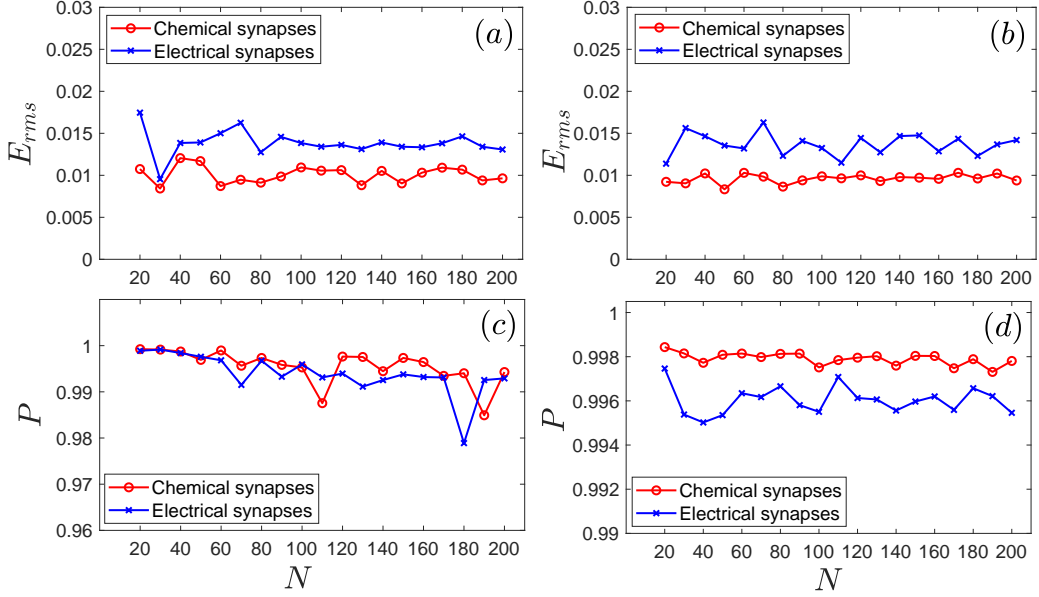}
\caption{\label{fig:epsart}
 {Dependence of reconstruction precisions on the network sizes for electrical and chemical synapses. Plots of $E_{rms}$ vs. $N$ and $P$ vs. $N$ in the fully connected networks (a)(c) and the sparse networks (b)(d). $m=10^5$, $K'=140 mV$. }}
\end{figure}

Next we still study the network of Fig. 1 with dynamic synapses.
Tsodyks and Markram (TM) model \cite{27,28} describes short-term
synaptic plasticity and a modified TM model is expressed by the
following form
\begin{eqnarray}
\dot{y}_{ij}(t) &=& \frac{1-y_{ij}}{\tau_{d}}-u_{ij}y_{ij}\sum _{k=1} ^{n}\delta (t-t_{j}^{(k)}), \\
I_{ij}^{(syn)}&=&\gamma_{ij}u_{ij}y_{ij}\frac{V_{rev}-V_{i}}{1+\exp(-(V_{j}-V_{th})/
\sigma)}
\end{eqnarray}
where $y_{ij}$ and $u_{ij}$ are two normalized variables. $y_{ij}$
indicate the short-term depression effect and present the
fraction of resources that remain available after neurotransmitter
depletion. $u_{ij}$ represent the short-term facilitation effect,
denoting the fraction of available resources ready for use
(release probability). In this Letter, we take the variables $y_{ij}
\in [0, 1]$ and the constants $u_{ij}$ distributed uniformly in $[0,
1]$. $\tau_{d}=1000 \ ms$ is a time constant. $\gamma_{ij}$ represent
the strength of chemical synapses. According to the statistical
curves of the coupling functions, we assume
\begin{equation}
\bar{I}_{ij}^{(syn)}=\hat{g}_{ij}V_{j}+\hat{\gamma}_{ij}\frac{1}{1+\exp(-(V_{j}-V_{th})/
\sigma)}.
\end{equation}
where the first and the second terms of the right hand side are
the electrical and chemical synapses, respectively. We take the
parameters same as in Figs. 3(a) and 3(b) and compute the numerical values
$\hat{g}_{i1}$ and $\hat{\gamma}_{i1}$ via Eq. (10). In Fig. 5 we
compare the numerical results $\hat{g}_{i1}$ and $
\frac{\hat{\gamma}_{i1}}{\langle V_{rev}-V_{i}\rangle}$ with the
actual parameters $g_{i1}$ and $\gamma_{i1}u_{i1}\langle y_{i1}\rangle$. The results display a good agreement for kinetic
synapses.

\begin{figure}[H]
\includegraphics[width=14cm]{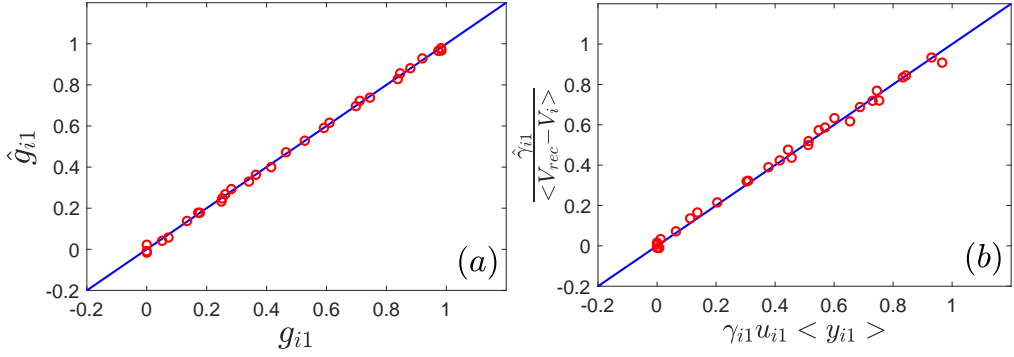}
\caption{\label{fig:epsart} Reconstruction of network of Fig.1
with dynamic synapses. (a) Numerical $\hat{g}_{i1}$ vs. actual
$g_{i1}$ (b)$ \frac{\hat{\gamma}_{i1}}{\langle
V_{rev}-V_{i}\rangle}$ vs. $\gamma_{i1}u_{i1}\langle
y_{i1}\rangle$. Sampling number $m=10^{5}$.}
\end{figure}

Now we discuss the effect of $\tau_{reset}$ on the results. Figures 6(a) and 6(b) show plots of
$E_{rms}$ vs. $\tau_{reset}$ in the static and dynamical synapses.
We obtain that the errors for the electric and chemical synapses are almost
independent of $\tau_{reset}$ except that the errors for the dynamical chemical
synapses in Fig. 6(b) are larger when $\tau_{reset}< 100 \ ms$. This shows that with a small resetting interval the reconstruction errors for detecting TM model increase
since over fast resetting results in the deviation of hidden variables $y_{ij}$: $\langle y_{ij}\rangle $ cannot return to the normal ranges for small $\tau_{reset}$.

\begin{figure}[H]
\includegraphics[width=14cm]{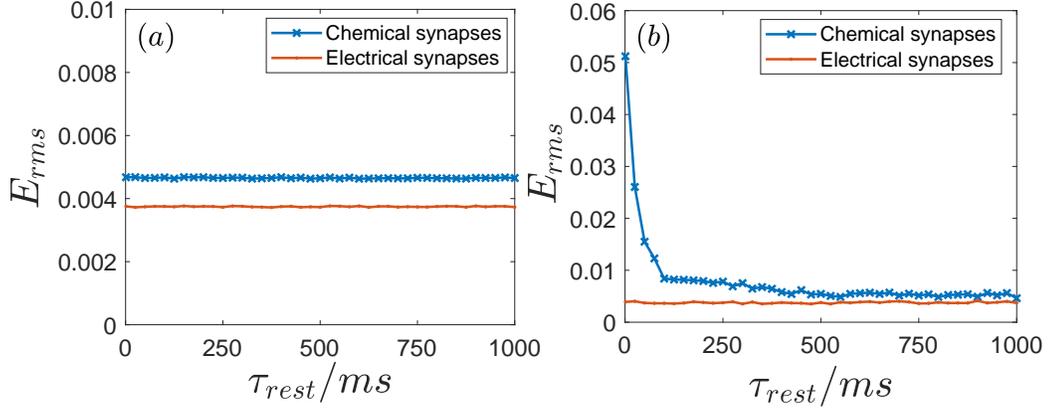}
\caption{\label{fig:epsart} Plots of errors $E_{rms}$ vs $\tau_{reset}$ for detecting network of Fig. 1 with the static synapses (a) and with the dynamical synapses (b).}
\end{figure}

In practical applications, random current resetting, i.e., resetting the injected current $I_{inject,j}$, is easily implemented.
Now we apply our findings in the network of Fig.1 by resetting injection current $I_{inject,j}=\frac{K_{j}}{\tau_{I}}$,
where $K_{j}$ and $\tau_{I}$ is the amplitude and the duration of random current resetting.
$K_{j}$  {is distributed} uniformly in $U(0,K)$.

Below we discuss the effect of the strength $K$ and the duration $\tau_{I}$ on reconstruction.
Simulation results are plotted in Fig.7. For $K=100\ \mu C$, $\tau_{I}=10 \ ms$, we compare the computed $\hat{g}_{i1}$ and $\frac{\hat{\gamma}_{i1}}{\langle V_{rev}-V_{i}\rangle}$ with the actual $g_{i1}$ and $\gamma_{i1}u_{i1}\langle y_{i1}\rangle$ in Fig. 7(a) and 7(b), respectively.
The computed errors of $\hat{g}_{i1}$ via Eq.(15) are plotted in Fig.7(c) (with $K=100\ \mu C$) and Fig.7(e) (with $\tau_{I}=10 \ ms$).
The computed errors of $\frac{\hat{\gamma}_{i1}}{\langle V_{rev}-V_{i}\rangle}$ are plotted in Fig. 7(d) (with $K=100\ \mu C$) and Fig. 7(f) (with $\tau_{I}=10 \ ms$).
Simulation results show that: for too large $\tau_{I}\geq 12 \ ms$  and too small $K \leq 90\ \mu C$ random current resetting cannot accurately estimate the links since over large $\tau_{I}$ and over small $K$ make the correlation between $r_{ij}$ and $x_j$ increase, i.e., Eq.(7) has a bias, $\langle r_{ij}F(x_{j})\rangle \neq \langle r_{ij}\rangle \langle F(x_{j})\rangle$ ($\langle r_{ij}F(x_{j})\rangle \neq 0$).
Compared with electric synapses, detection of chemical synapses shows better results since the nonlinear coupling of chemical synapses weakens this correlation between $r_{ij}$ and $x_j$.
Comparing random variable (membrane potential) resetting with random current resetting, we observe that random variable resetting has a more satisfactory reconstruction since random current resetting indirectly changes variable $V_j$ whereas random variable resetting directly changes variable $V_j$.
It takes some time for the current resetting to change the membrane potential, which results in a correlation between the reset membrane potential and other variables.
The larger the correlation between the membrane potentials after resetting, the less accurate the reconstruction results will be.

\setlength\parskip{.0\baselineskip}

\begin{figure}[H]
\includegraphics[width=14cm]{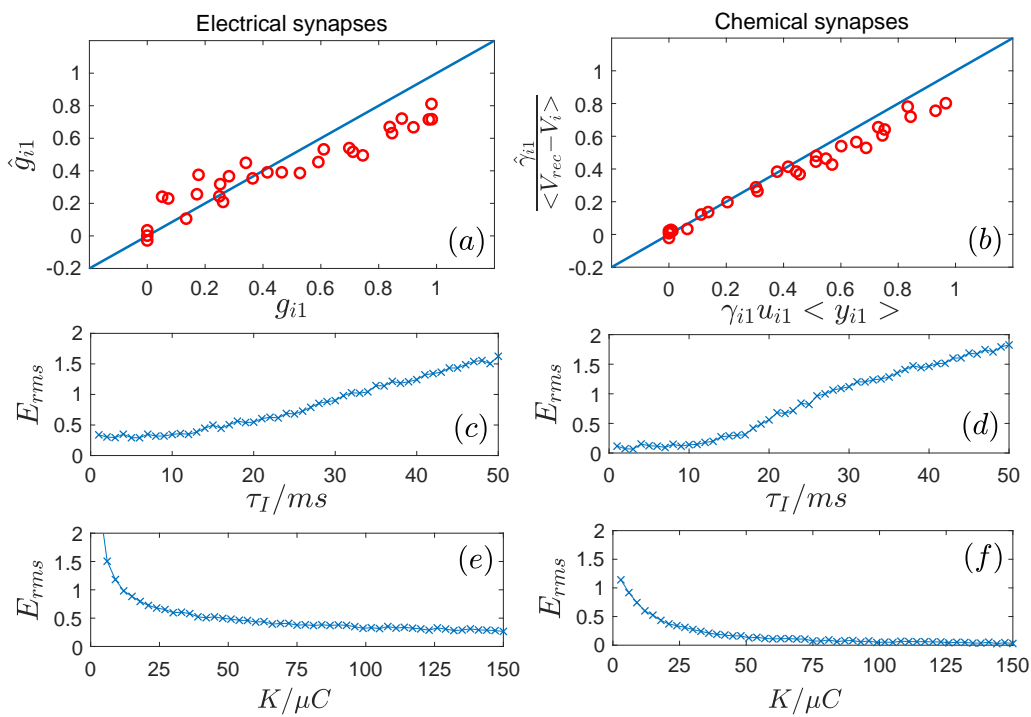}
\caption{\label{fig:epsart} Simulation results of network of Fig.1 by random current resetting.
(a) Numerical results $\hat{g}_{i1}$ plotted against the actual coefficients $g_{i1}$.
(b) Numerical results $\frac{\hat{\gamma}_{i1}}{\langle V_{rev}-V_{i}\rangle}$ plotted against the actual coefficients $\gamma_{i1}$.
The average reconstruction errors of $\hat{g}_{i1}$ at $K=100 \ \mu C$ (c) and at $\tau_{I}=10 \ ms$ (e). The average errors of $\frac{\hat{\gamma}_{i1}}{\langle V_{rev}-V_{i}\rangle}$ at $K=100 \ \mu C$ (d) and at $\tau_{I}=10 \ ms$ (f).
}
\end{figure}

Random variable resetting is one of the external control methods implemented in some real systems, which are invasive, and to some extent, active. Scientific tests have found that the reinitiation of epileptiform activity follows stimulus removal immediately or after a few minutes \cite{30,31,32}. We expect the method of detecting directed interaction functions of a general dynamic network by random variable resetting has potential application in real networks.

In conclusion, we propose a novel method of reconstructing the directed interaction functions of a general dynamical network.
Through reinitializing the state variable of the target node, the interaction functions of the target node to the response ones can be directly inferred via analyzing the statistical characteristic of dynamics.
The statistical characteristic through random variable resetting makes the dynamics of the response nodes reduced to the equivalent coupling functions and their fluctuations,
furthermore the equivalent coupling functions imply the interactions of the target node to the response ones.
This method is applicable to a wide class of directed networks with nonlinear dynamics and strong noise, while avoiding the needs of the appropriate test function and transformation to phases \cite{25}.
Especially this method can be applicable to the controllable networks with the hidden variables in local dynamics and interaction functions.


\end{document}